\documentclass[twocolumn]{aastex63}

\received{October 22, 2019}
\revised{December 12, 2019}
\accepted{December 29, 2019}
\submitjournal{ApJ}

\shorttitle{Chromospheric heating II}
\shortauthors{Abbasvand et al.}

\begin{document}

\title{Chromospheric Heating by Acoustic Waves Compared to Radiative Cooling:\\
II -- Revised Grid of Models}

\correspondingauthor{Michal Sobotka}
\email{michal.sobotka@asu.cas.cz}


\author[0000-0002-7912-8087]{Vahid Abbasvand}
\affiliation{Astronomical Institute of the Czech Academy of Sciences (v.v.i.) \\
Fri{\v c}ova 298, 25165 Ond{\v r}ejov, Czech Republic}
\affiliation{Astronomical Institute of Charles University, Faculty of Mathematics and Physics \\
V Hole\v{s}ovi\v{c}k\'{a}ch 2, 180 00 Praha 8, Czech Republic}

\author[0000-0001-5439-7822]{Michal Sobotka}
\affiliation{Astronomical Institute of the Czech Academy of Sciences (v.v.i.) \\
Fri{\v c}ova 298, 25165 Ond{\v r}ejov, Czech Republic}

\author[0000-0002-5778-2600]{Petr Heinzel}
\affiliation{Astronomical Institute of the Czech Academy of Sciences (v.v.i.) \\
Fri{\v c}ova 298, 25165 Ond{\v r}ejov, Czech Republic}

\author[0000-0002-6345-1007]{Michal \v Svanda}
\affiliation{Astronomical Institute of the Czech Academy of Sciences (v.v.i.) \\
Fri{\v c}ova 298, 25165 Ond{\v r}ejov, Czech Republic}
\affiliation{Astronomical Institute of Charles University, Faculty of Mathematics and Physics \\
V Hole\v{s}ovi\v{c}k\'{a}ch 2, 180 00 Praha 8, Czech Republic}

\author[0000-0002-9220-4115]{Jan Jur{\v c}{\'a}k}
\affiliation{Astronomical Institute of the Czech Academy of Sciences (v.v.i.) \\
Fri{\v c}ova 298, 25165 Ond{\v r}ejov, Czech Republic}

\author[0000-0003-2500-5054]{Dario del Moro}
\affiliation{Department of Physics, University of Roma Tor Vergata \\
Via della Ricerca Scientifica 1, I-00133 Rome, Italy}

\author[0000-0002-2276-3733]{Francesco Berrilli}
\affiliation{Department of Physics, University of Roma Tor Vergata \\
Via della Ricerca Scientifica 1, I-00133 Rome, Italy}

\begin{abstract}

Acoustic and magnetoacoustic waves are considered to be possible agents of chromospheric
heating. We present a comparison of deposited acoustic energy flux with total integrated
radiative losses in the middle chromosphere of the quiet Sun and a weak plage.
The comparison is based on a consistent set of high-resolution observations acquired
by the IBIS instrument in the \ion{Ca}{2} 854.2~nm line. The deposited acoustic-flux
energy is derived from Doppler velocities observed in the line core and a set of 1737
non-LTE 1D hydrostatic semi-empirical models, which also provide the radiative losses.
The models are obtained by scaling the temperature and column mass of five initial
models VAL B--F to get the best fit of synthetic to observed profiles. We find that
the deposited acoustic-flux energy in the quiet-Sun chromosphere balances
30--50 \% of the energy released by radiation. In the plage, it contributes by 50--60 \%
in locations with vertical magnetic field and 70--90 \% in regions where the magnetic
field is inclined more than 50\degr ~to the solar surface normal.

\end{abstract}


\section{Introduction} \label{sec:intro}

The energy released by radiation from the solar chromosphere is mostly concentrated in the strong
lines of \ion{Ca} {2}, \ion{Mg}{2}, and hydrogen Lyman-$\alpha$ as well as in the H$^-$ continuum
\citep{Vernazza81}.
It is characterized by net radiative cooling rates (radiative losses). The estimate of the total radiative losses
integrated over the height of the chromosphere is 4300~W~m$^{-2}$ in the quiet Sun \citep{Avrett81}
and it is by a factor of 2 to 4 higher in active regions \citep{Withbroe77}. This radiative cooling must
be balanced by heating processes that deliver energy to the chromosphere.
The dominant process heating the upper layers of the solar atmosphere is still being intensively
debated. There are many agents considered by various studies \citep[see][for a review]{Jess15}, which
could be grouped into two fundamentally different classes: (1) Mechanisms that are usually connected
with heat releases during the reconnection processes in the magnetic field and (2) mechanisms that are
associated with the deposit of energy by various kinds of waves. In the present work, we focus on
the latter class.

Waves are generated near the solar surface by turbulent motions of the convective plasma
and propagate upwards, where they can dissipate a considerable part of their energy in the
chromosphere. In the magnetized solar atmosphere, there exist three principal types of MHD waves:
Alfv\'en waves as well as fast and slow magnetoacoustic waves \citep{Khomenko09,KhomenkoCally12}. 
These waves with magnetic component are considered extremely important.
Low-frequency waves may represent a significant source of energy, related to the so-called
magnetic portals \citep{Jefferies06} or to magnetoacoustic-gravity waves \citep{Jefferies19}.
These waves are generally not allowed to
propagate higher into the atmosphere because their frequency does not exceed the expected
photospheric cutoff frequency 5.2 mHz \citep{BelLeroy77}. However, in regions where the
photospheric magnetic field is inclined with respect to the gravity vector, the cutoff
frequency can be lowered by means of the ramp effect \citep{Stangalini11}.
This allows waves with frequencies far below 5.2 mHz, which would otherwise be trapped in the
photosphere, to propagate into the upper atmosphere. Recently, \citet{Rajaguru19} used
{\em Solar Dynamics Observatory} \citep{Pesnell12} data to discuss thoroughly the relations 
between magnetic field properties and the propagation of acoustic waves.

\defcitealias{Sobotka16}{Paper~I}

\citet[hereafter Paper~I]{Sobotka16} studied the hypothesis of chromospheric heating
by acoustic and magnetoacoustic waves.
This work was based on 70-minute long observations of the \ion{Ca}{2} 854.2 nm
near-infrared line with the {\em Interferometric Bidimensional Spectrometer} \citep[IBIS,][]
{Cavallini06} at the {\em Dunn Solar Telescope}. Sequences of Dopplergrams were used to compute
energy flux carried by (magneto)acoustic waves and a simplistic grid of seven semi-empirical
non-LTE (that is, with departures from the local thermodynamic equilibrium, LTE) hydrostatic models
of the atmosphere to estimate the radiative losses.
Although the propagation and dissipation of waves in the chromosphere is generally a time-dependent
process, 1D static models are still a reasonable tool to represent time-averaged physical conditions in
long-lived structures of the solar atmosphere \citep{Heinzel19}. In \citetalias{Sobotka16} we
found that a significant portion of the radiative losses could be replenished by the dissipative acoustic
flux generated by the $p$-modes, which were converted to magnetoacoustic modes in the inclined
magnetic field. We have shown that there was a correlation between the estimate of the acoustic
flux dissipation in the region and the estimate of the radiative losses. That provided an indication that
there indeed was a contribution of the acoustic waves to the heating of the chromosphere, which
should be considered.

A weak point of that work was a scarce grid of only seven atmospheric models, which were
used to calculate the radiative cooling rates as well as the acoustic fluxes.
In this paper we present an analysis of the same observations as in \citetalias{Sobotka16}, applying
a much finer grid of models obtained by up- and downscaling the frequently used semi-empirical
models of \citet[][VAL~models]{Vernazza81}, to make sure that the comparison of deposited
acoustic flux and radiative losses is more conclusive.

\section{Observations and data analysis} \label{sec:obs}

The target, a slowly decaying bipolar active region NOAA 11005 located at 25.2 N and 10.0 W
(heliocentric angle $\vartheta = 23$\degr) was observed with IBIS on 2008 October 15 from
16:34 to 17:44 UT. A pore surrounded by a superpenumbra \citep{Sobotka13} was present
in the leading part of the region. A weak chromospheric plage with equal polarity was located
near the pore (see Figure~1 in \citetalias{Sobotka16}). The IBIS data set was described in detail
in \citet{Sobotka12}, \citet{Sobotka13}, and \citetalias{Sobotka16}. Two spectral lines,
\ion{Ca}{2} 854.2~nm (intensity only) and \ion{Fe}{1} 617.33~nm (full Stokes vector) were
observed simultaneously. The relevant parameters for both spectral lines are summarized in
Table \ref{tab:obs}.

%
\begin{table}[]
\caption{Parameters of the IBIS data set}
\label{tab:obs}
\centering
\begin{tabular}{ll}
\hline \hline
\noalign{\smallskip}
    Date and time                  & 2008 October 15, 16:34 -- 17:44 UT \\
    Scanned lines                  & \ion{Fe}{1} 617.33~nm ($I, Q, U, V$); \\
                                         & \ion{Ca}{2} 854.2~nm ($I$) \\
    No. of sp. points             & 21 in each line \\
    Wavelength spacing         & 2 pm (\ion{Fe}{1}); 6 pm (\ion{Ca}{2}) \\
    Field of view                   & 38\farcs 0 $\times$ 71\farcs 5; part of  NOAA 11005 \\
    Region of interest            & 13\farcs 7 $\times$ 39\farcs 7 \\
    Pixel size                        & 0\farcs 167 $\times$ 0\farcs 167 \\
    Spatial resolution            & 0\farcs 4 \citep{Sobotka12} \\
    Exposure time                & 80 ms \\
    Spectral scan time          & 52 s (time resolution) \\
    No. of sp. scans             & 80 \\
\hline
\end{tabular}
\end{table}
%

The data-processing procedures were described thoroughly by \citet{Sobotka12} and \citet{Sobotka13}.
We concentrated on the region of interest (ROI) 13\farcs7 $\times$ 39\farcs7 (82 $\times$ 238 pixels)
that included the plage eastward of the pore and a quiet-Sun region.
The location of ROI in the whole field of view is shown in Figure~1 of \citetalias{Sobotka16}.
The following observables were obtained from the IBIS data set (see \citetalias{Sobotka16} for details):

(1) The magnetic-field vector in the photosphere, retrieved from the full-Stokes spectral scan of
\ion{Fe}{1} 617.33~nm taken at 17:10 UT by means of the Stokes inversion code based on
response functions \citep[SIR,][]{RuizCobo92}. 
Because the acoustic flux in the chromosphere
is very sensitive to the inclination of magnetic field to the solar surface normal, we revise the
calculations of \citetalias{Sobotka16} to avoid uncertainties in regions with weak fields, where the
Stokes $Q$, $U$, and $V$ signals are dominated by noise. In such cases, the code returns an
unreliable line-of-sight (LOS) inclination near 90\degr ~(transversal direction), which can be removed
by a mask that for the fields weaker than 350 G sets to zero all inclination angles in the range
$\pm 5$\degr ~from the transversal direction. We update the method of removing the 180\degr
~ambiguity of the LOS azimuth, using the AMBIG code \citep{Leka09}. In the revised calculations,
we also removed a bug in the transformation from the LOS reference frame to the local reference
frame (LRF) that led to an overestimation of magnetic inclination in the plage. We use routines from
the AZAM code \citep{Lites95} for the transformation from the LOS reference frame to LRF.

(2) Time-dependent line-of-sight velocities, measured in the inner wings of the
\ion{Ca}{2} 854.2~nm core at $\pm 18$~pm and in the line center. According to
\citet{Cauzzi08}, the inner wings are formed at approximately 900--1000~km above the optical depth
$\tau_{500} = 1$ and the line centre at 1400--1500 km. These velocities are used to calculate
power spectra of oscillations in two different layers of the chromosphere.

(3) Mean profiles of the \ion{Ca}{2} 854.2~nm line, obtained by time-averaging over the
70-minute observing period of the observed profiles with removed Doppler shifts.
They are used to find the most appropriate semi-empirical models at each location in ROI.

The cadence of scans (52 s) and the length of the time-series sets the maximum detectable
frequency of oscillations to 9.6~mHz and the frequency resolution to 0.24~mHz, suitable to
analyze low-frequency waves. Power spectra of the line-of-sight velocities were calculated
using the standard Fourier analysis and the results were discussed in \citet{Sobotka13}.
The acoustic energy fluxes at the heights of
900 km and 1500 km were estimated following \citet{Bello09}. The method consists in
an integration over frequencies of the product of gas density, spectral power density, and
group velocity of energy transport. The frequency integration range spans between the
acoustic cutoff frequency and the maximum detected frequency---see \citetalias{Sobotka16}
for details. The quantities depend on the gas density and pressure at the given height, which are
taken from the model atmosphere. The acoustic cutoff frequency is proportional to the cosine
of the magnetic field inclination to the solar surface normal.

Taking into account that the acoustic flux estimated at 900 km approximates the incoming
energy flux that may partly dissipate in the atmosphere, while the acoustic flux at 1500 km
corresponds to the energy flux that has passed without dissipation, then the deposited
acoustic flux in the chromospheric layers between 900 km and 1500 km is
the difference between these two fluxes. The corresponding net radiative cooling rates
are integrated in the same range of heights.

\section{Grid of chromospheric models} \label{sec:grid}

The grid of models has been constructed keeping the number of free parameters as low
as possible. We scale the existing set of semi-empirical 1D hydrostatic models VAL A--F,
which describe the solar atmosphere from intranetwork to bright network features, by
changing independently the temperature and column mass stratifications. Adopting
the column mass as an independent variable makes it possible to conserve the condition
of hydrostatic equilibrium in the scaled models.

%
   \begin{figure}[t]
       \centering
\includegraphics[width=0.45\textwidth]{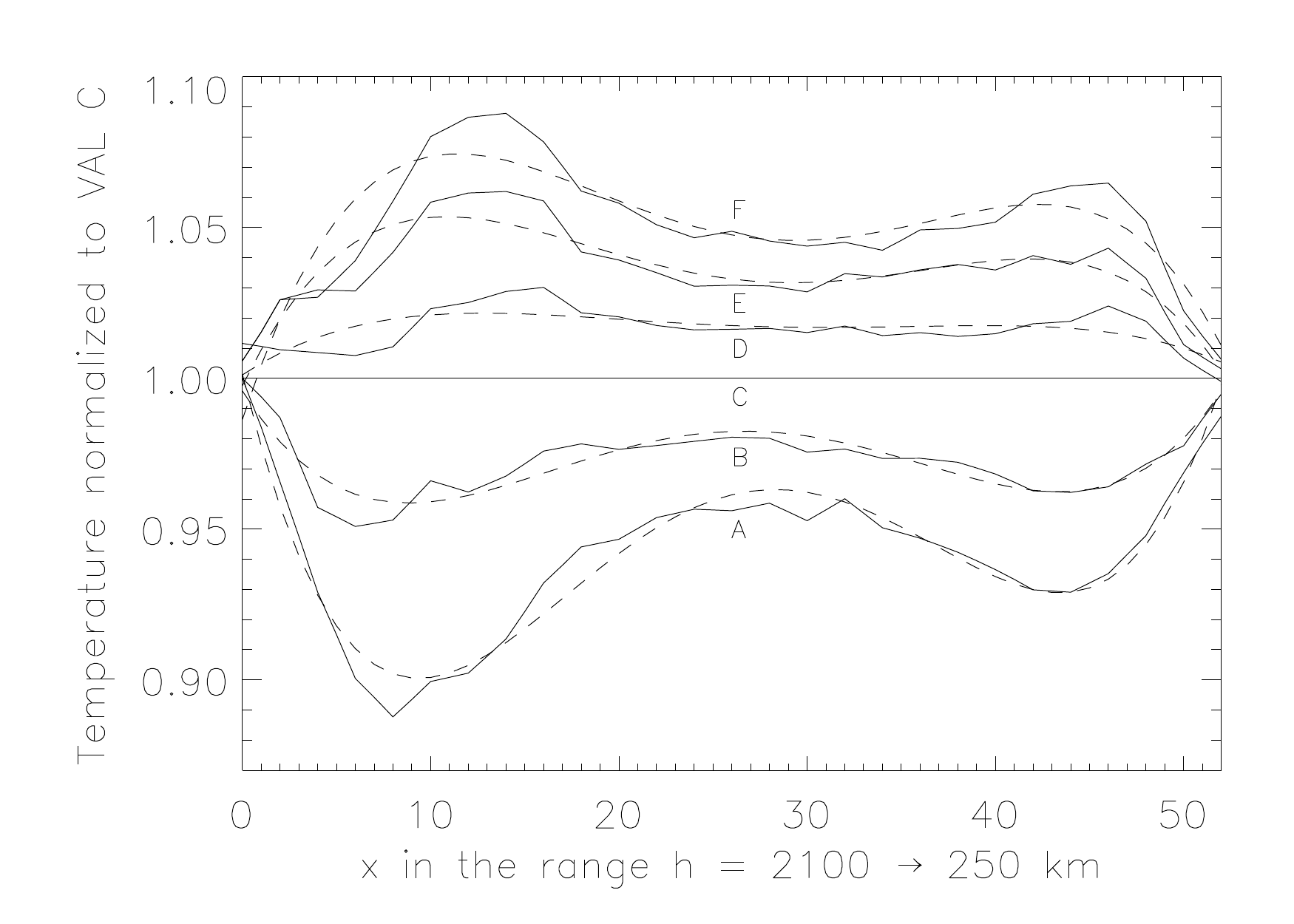}
\caption{Temperature stratifications of models VAL A--F normalized to VAL C
            (solid lines) in the range $h$ from 2100~km to 250~km.
            Dashed lines show the approximation by fourth-order polynomials.}
       \label{Fig:poly}
   \end{figure}
%

The original VAL models specify the column mass $m$, optical depth $\tau_{500}$
at 500 nm, temperature $T$, microturbulent velocity $v_{\rm t}$, hydrogen density $n_{\rm H}$,
electron density $n_{\rm e}$, total pressure $P_{\rm tot}$, gas pressure to total pressure 
ratio $P_{\rm g}/P_{\rm tot}$, and density $\rho$ in 52 geometrical heights $h$ ranging
from $-75$ km to 2400--2700 km ($h = 0$ is at $\tau_{500} = 1$). The microturbulence
stratification is practically equal in all models A--F at the heights from $-75$~km to 2000~km,
that is, in the whole photosphere and chromosphere, and it is not changed in scaled models. 
To improve numerical accuracy of the following computations, we resampled the models
to 103 heights by means of the linear interpolation. The scaling of these initial models
is done in two steps: (1) $m$ and $T$ stratifications are changed and (2) the other
model quantities are re-calculated.

%
   \begin{figure*}[t]
       \centering
\includegraphics[width=0.8\textwidth]{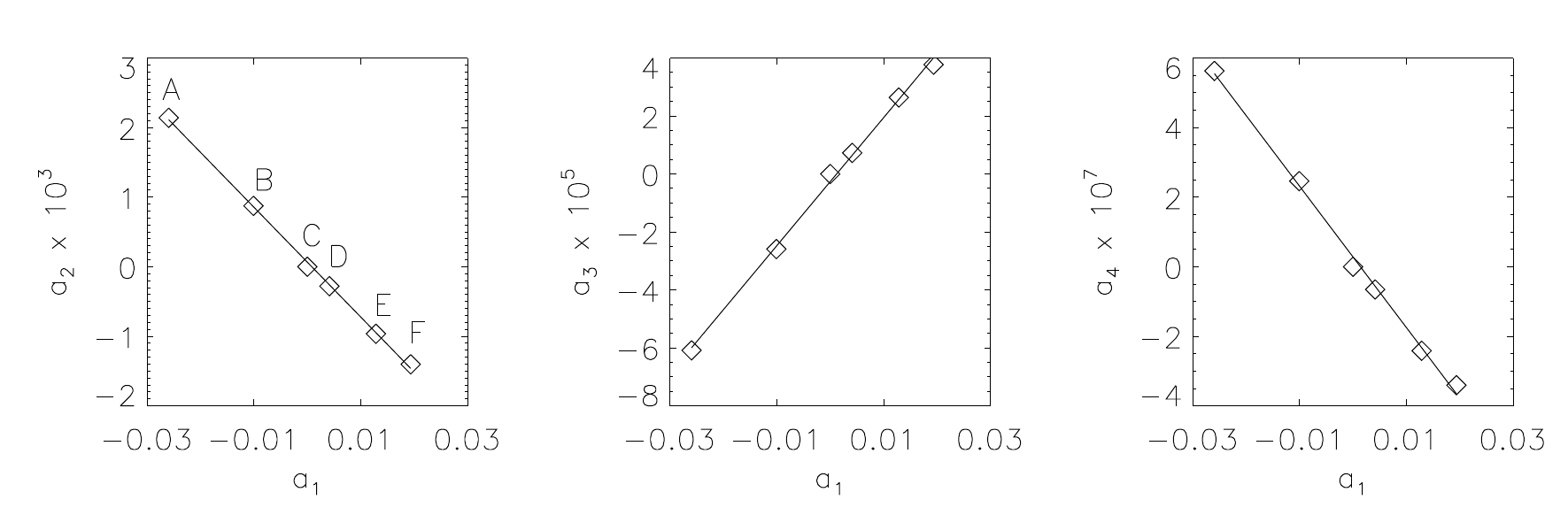}
\caption{Linear relations of the fourth-order polynomial coefficients $a_{1-4}$
             of the temperature scaling. Diamonds show values corresponding
             to the models VAL A--F. }
       \label{Fig:a2-4}
   \end{figure*}
%

We can find an adequate way to change $m$ and $T$ by adopting, for example,
the VAL C model as the initial one and reproducing the stratifications of all the other
VAL models. In general, the scaling algorithm must be able to reproduce all the
stratifications using any VAL model as the initial one. The following solution was adopted:
The initial column mass stratification $m_0(h)$ is changed for $h \geq 450$ km
($m < 0.1$ g cm$^{-2}$) whereas it is kept unchanged in deeper layers.
The changes are controlled by scaling parameter $p_{\rm m}$, so that the scaled column mass is
\begin{eqnarray*}
m_x &=& b_x m_{0x}, \ {\rm where} \\
b_x &=& p_{\rm m} \  {\rm for} \ h \geq 2000 \, {\rm km}, \\
b_x &=& p_{\rm m} + (1-p_{\rm m})\, x/(x_{450}-x_{2000}) \\
& & {\rm for} \  2000 \, {\rm km} > h \geq 450 \, {\rm km},
\end{eqnarray*}
and $x$ is the sampling (row) index in the model table.
In upper layers, from the maximum height down to $h \approx 2000$~km,
$b_x$ is simply a multiplicative factor equal to $p_{\rm m}$.  In deeper layers
($h < 2000$ km), this factor linearly decreases or increases to reach unity at
$h \approx 450$ km.

The initial temperature stratification is changed in the range of heights from
approximately 250~km to 2100~km; the upper and lower parts remain equal to
the initial model. To derive the scaling parameter $p_{\rm T}$, we normalize
the temperature stratifications $T_x$ of all VAL models to VAL C:
$A_x = T_x /T_{x \, {\rm VAL\, C}})$. These normalized stratifications
can be approximated by a fourth-order polynomial in the form
$A^\prime (x) = 1 + \sum_{i=1}^{4} a_i x^i$  (Figure~\ref{Fig:poly}).
The coefficients $a_i$ are mutually dependent and $a_{2-4}$ can be
expressed by multiples of $a_1$ (Figure~\ref{Fig:a2-4}), particularly,
$a_{i = 2,3,4} = a_1 q_i$, where $q_i = [-7.90 \cdot 10^{-2}, 2.21 \cdot 10^{-3}, -2.03 \cdot 10^{-5}]$.
The values of $q_i$ are practically independent of the initial model.
Thus, $a_1$ can be used as the temperature scaling parameter,  $p_{\rm T} = a_1$.
The scaled temperature stratification (in the range 250--2100 km) is then calculated as
\[ T_x = T_{0x}  (1 + p_{\rm T} x + p_{\rm T} \sum_{i=2}^{4} q_i x^i), \]
where $T_{0x}$ are the initial temperatures.

%
   \begin{figure}
       \centering
\includegraphics[width=0.45\textwidth]{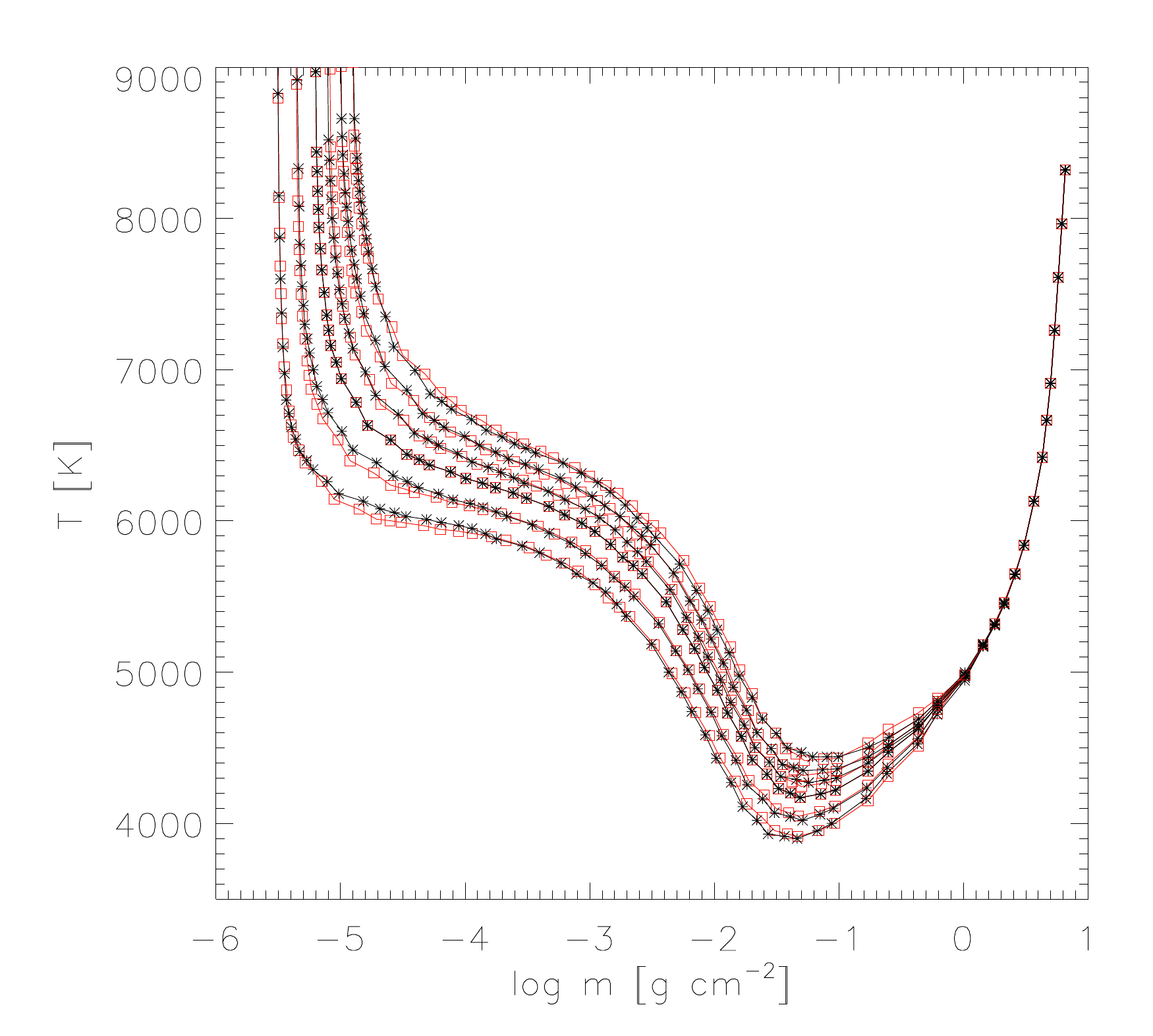}
\caption{Temperature versus column mass plots of the VAL A--F models scaled from
             the model C. Red and black lines show the scaling results and the original
             VAL models, respectively. The symbols denote sampling of the models.}
       \label{Fig:Cscaled}
   \end{figure}
%

For each of the initial models VAL we calculated a grid of 2806 scaled models using
a combination of the parameters $p_{\rm m} = [0.5,\, 0.6,\, ...,\, 4.9,\, 5.0]$ and
$p_{\rm T} = [-0.030,\, -0.029,\, ...,\, 0.029,\, 0.030]$. In Figure~\ref{Fig:Cscaled}
we show that the models VAL A--F can be reproduced by scaling the model C, applying
the sets of parameters  $p_{\rm m} = [0.5,\, 0.7,\, 1.0,\, 1.3,\, 1.6,\, 2.0]$ and
$p_{\rm T} = [-0.021,\, -0.010,\, 0.000,\, 0.006,\, 0.012,\, 0.020]$, respectively.
In total, we have 16836 models in the grid parametrized by $p_{\rm m}$,
$p_{\rm T}$, and the initial model selection.

%
   \begin{figure*}[t]
       \centering
\includegraphics[width=1.0\textwidth]{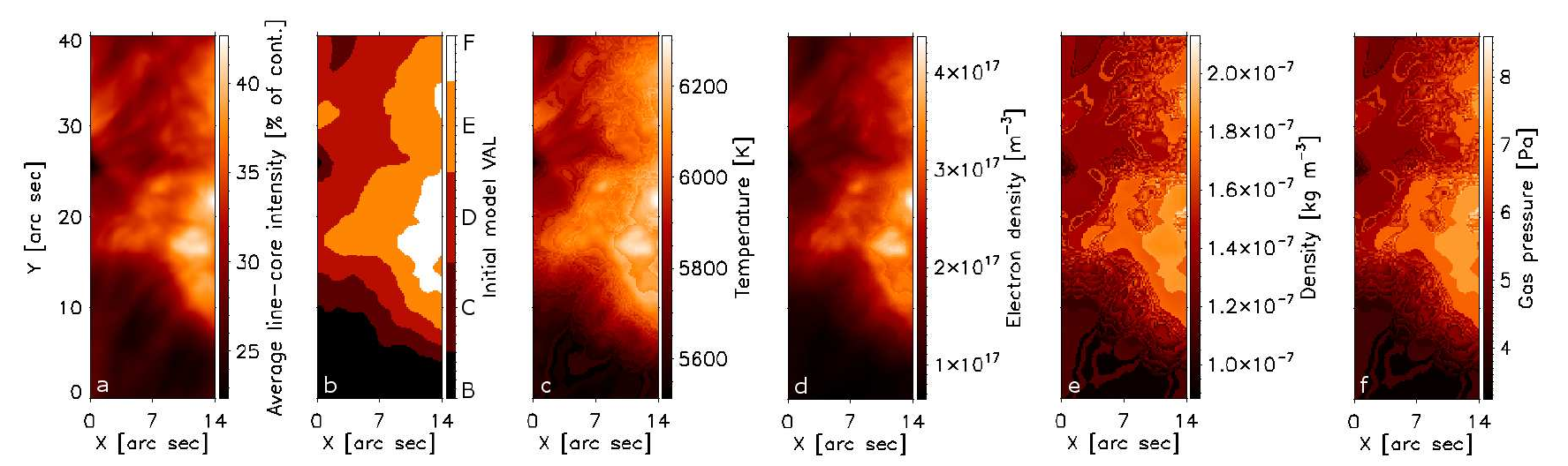}
\caption{Maps of (a) \ion{Ca}{2} 854.2~nm line-core intensity, (b) selection of initial models,
(c) temperature, (d) electron density, (e) density, and (f) gas pressure. The maps (c--f) are
retrieved from resulting scaled models at $h = 900$ km.}
       \label{Fig:modmaps}
   \end{figure*}
%

The models, which have been obtained by varying the temperature structure on the column-mass
scale, are used as the input atmospheres in the non-LTE  radiative-transfer codes MALI.
MALI stands for the Multi-level Accelerated Lambda Iterations technique with preconditioning
of the statistical-equlibrium equations according to \citet{Rybicki91,Rybicki92}. The density
structure is obtained from the hydrostatic equilibrium for a given temperature structure and the
radiative transfer equation is solved in a semi-infinite atmosphere subject to standard boundary
conditions. First we solve the hydrogen problem using a 5-level plus continuum atomic model.
To obtain the gas density, we add the helium but we neglect the helium ionization in computing
the electron density within the atmosphere. For hydrogen resonance lines Lyman-$\alpha$ and
Lyman-$\beta$ we consider the standard angle-averaged partial frequency redistribution (PRD)
for the scattering part of the source function and all other lines are treated in the complete
redistribution.

The microturbulence, taken from the initial VAL model, is consistently included in the hydrostatic
equilibrium as the turbulent pressure and it enters also in the line broadening calculations. The resulting
electron densities are then used, together with the temperature structure, in the \ion{Ca}{2} version
of the MALI code, which solves the non-LTE problem for a 5-level plus continuum model of
\ion{Ca}{2}--\ion{Ca}{3} ions. This governs the \ion{Ca}{2} H and K resonance lines,
the \ion{Ca}{2} infrared triplet lines and the five continua. Again, PRD is used for both
resonance lines H and K. We perform detailed synthesis of the \ion{Ca}{2} 854.2 nm line,
using all relevant broadening mechanisms. Due to uncertainties in the  Van der Waals damping
parameter, we adjust it in order to better fit the wings of this line. From the grid of models
we compute the synthetic line profiles of the 854.2 nm line and use them to find the best fit
to the observed profiles (see Section~\ref{sec:comp}) at all positions in ROI. 

The energy losses due to radiation are characterized by the net radiative cooling rates or simply
the net radiative losses. For selected best-fit models, at each position in ROI and model height,
we finally compute the net radiative cooling rates due to hydrogen, \ion{Ca}{2}, and \ion{Mg}{2}.
These are the main contributors in the solar chromosphere as demonstrated in \citet{Vernazza81}.
The \ion{Mg}{2} version of our MALI code uses a 5-level plus continuum
\ion{Mg}{2}--\ion{Mg}{3} atomic model and treats the two resonance lines \ion{Mg}{2} h and k
and the triplet lines. Both resonance lines are again computed using the PRD approach.
Contrary to optically-thin losses from the solar corona, the chromospheric losses in the above
listed lines must be computed by solving the complex non-LTE radiative transfer problem,
because these lines are optically thick and there is no simple and sufficientlly precise prescription
for an easy estimate of such losses. The resulting net radiative losses are then integrated along
the atmospheric height in the range from 900~km to 1500~km, which corresponds to the range
where the measured acoustic energy flux is dissipated (see Section \ref{sec:obs}). At such
heights, the losses are dominated by \ion{Ca}{2} lines with some contribution of \ion{Mg}{2}
lines h and k and hydrogen continua. A strong coolant is the hydrogen Lyman-$\alpha$ line,
but this is formed higher in the atmosphere close to the transition zone and thus does not enter
our integration over the heights of interest.

\section{Deposited acoustic flux compared to radiative cooling} \label{sec:comp}

%
   \begin{figure*}[t]
       \centering
\plottwo{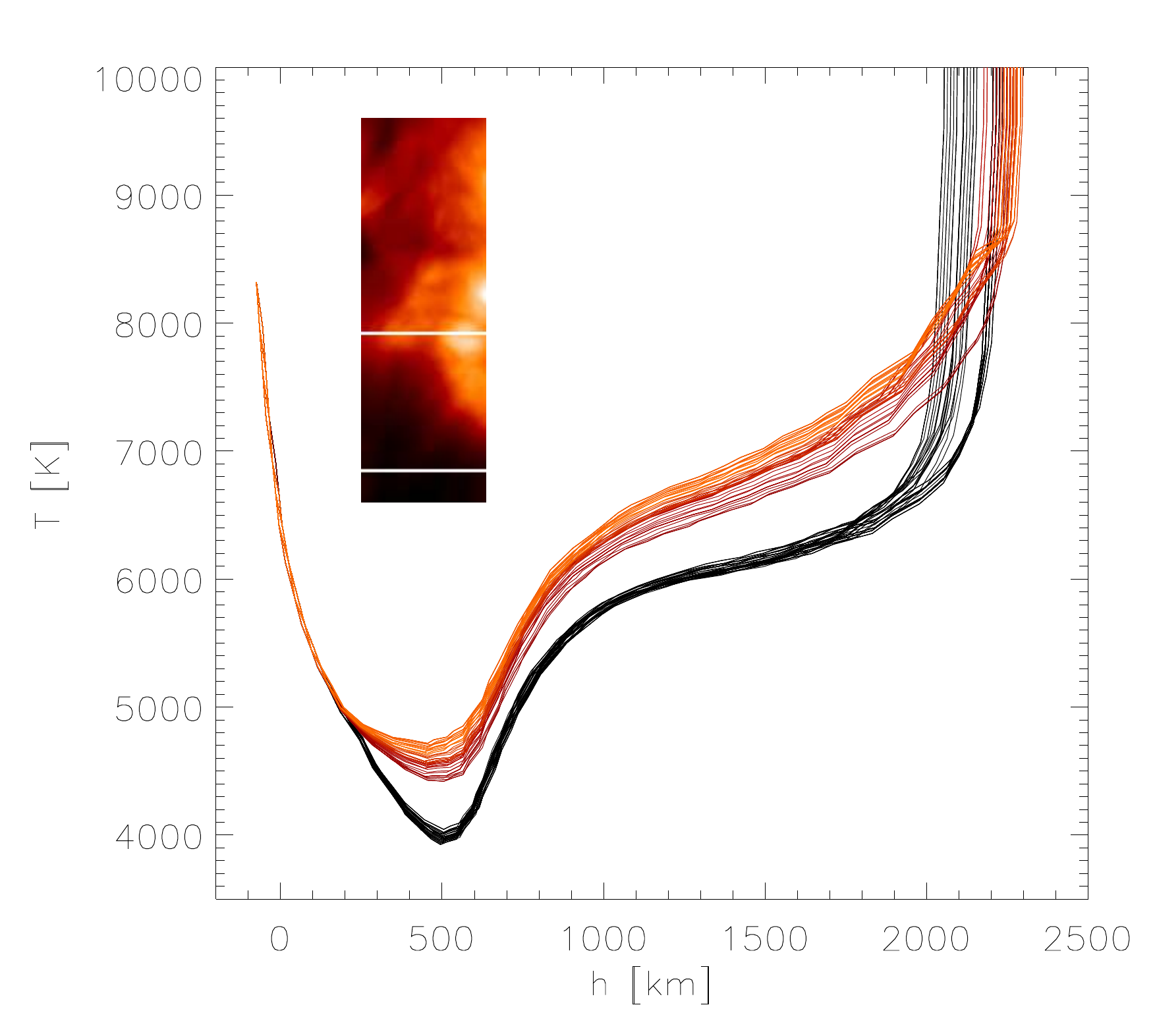}{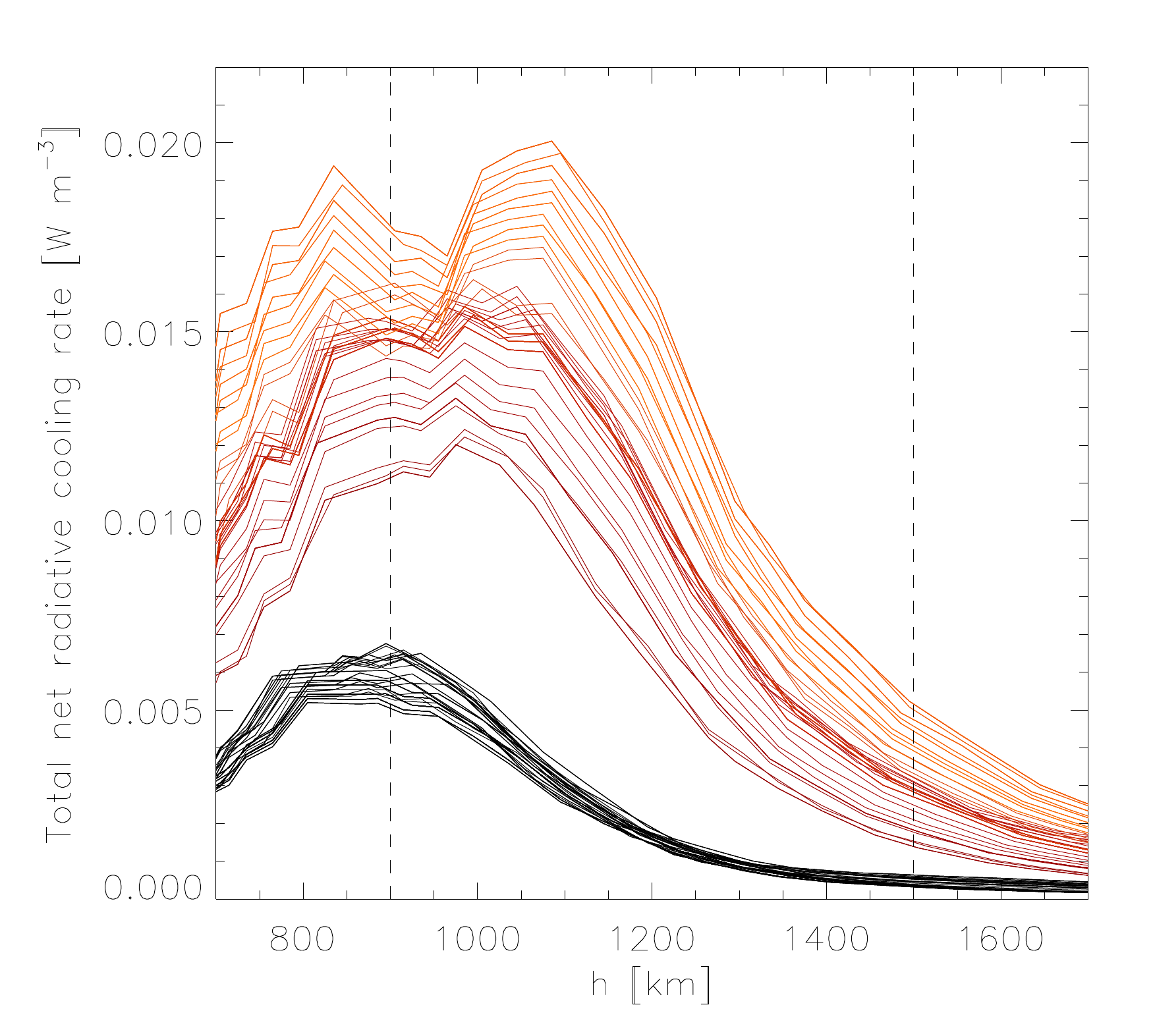}
\caption{Left: Examples of quiet-Sun temperature stratifications with height (black)
at positions along the bottom white line in the inset image of the line-core intensity.
Red-to-orange temperature curves correspond to positions from left to right along
the top white line that crosses the plage.
Right: Total net radiative cooling rates versus height in quiet Sun (black) and
plage (red--orange) at the same positions. Vertical dashed lines delimit the integration range.}
       \label{Fig:T-example}
   \end{figure*}
%

%
   \begin{figure*}
       \centering
\includegraphics[width=0.7\textwidth]{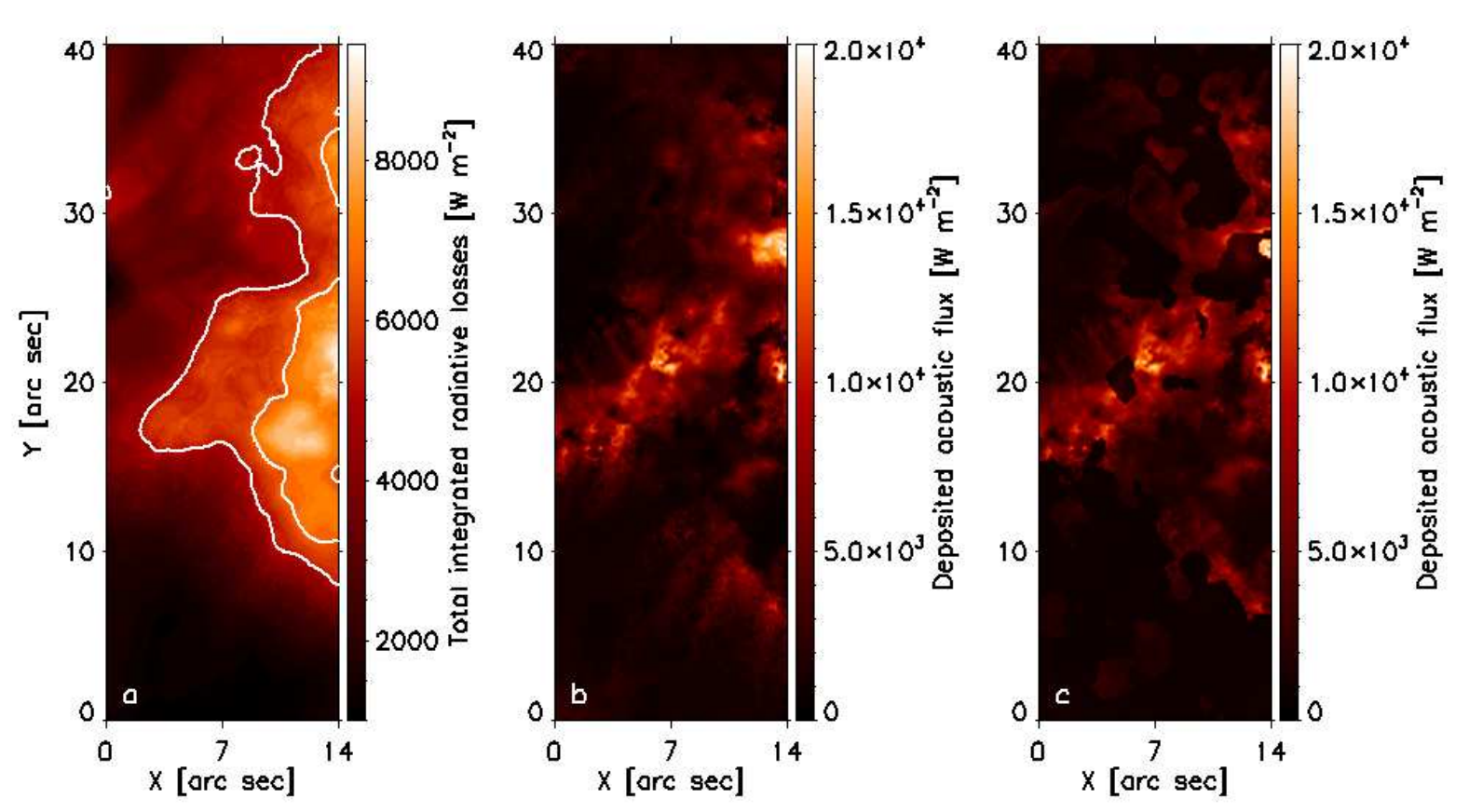}
\caption{(a) The total integrated radiative losses map with contours of 5000 and 
6500~W~m$^{-2}$. The deposited acoustic flux maps are calculated using (b) the original
magnetic inclination angles and (c) the corrected ones.}
       \label{Fig:rlosses}
   \end{figure*}
%

The initial models VAL B--F were assigned to different areas in ROI in accordance with
the brightness of the \ion{Ca}{2} 854.2 nm line core (Figures~\ref{Fig:modmaps}a and~b).
Then, a scaled model from the grid, which provided the best match of the synthetic
$I_\lambda^{\rm syn}$ to the local mean observed profile $I_\lambda^{\rm obs}$ of
the line, was assigned to each of 19516 positions in ROI. The models have been found
by minimizing the merit function
\[ \chi^2({\rm model}) = \sum_\lambda [I_\lambda^{\rm syn}({\rm model}) - I_\lambda^{\rm obs}]^2. \]
In total, 1737 different models were used.
The retrieved values of model $T$, $n_{\rm e}$, $\rho$, and $P_{\rm g}$ at the
geometrical height $h = 900$~km are shown in Figures~\ref{Fig:modmaps}c--f.
Because each position in ROI is characterized by a 1D model unrelated to its
surroundings, the maps of $\rho$, and $P_{\rm g}$ have a noisy appearance.
The local differences, however, are small---for example, the standard deviation
of gas pressure in the ``noisy'' subfield 6\arcsec$<x<$11\arcsec, 30\arcsec$<y<$40\arcsec
~is 0.5~Pa (9 \% of the mean value). Examples of typical temperature stratifications in
quiet-Sun and plage are depicted in the left panel of Figure~\ref{Fig:T-example}.
The inset image of the \ion{Ca}{2}~854.2~nm line core intensity shows positions
of the examples: the quiet-Sun temperatures (black curves) correspond to the bottom
white line, while those in the plage are taken along the top white line and the colors
of the temperature curves change from dark red to orange moving from left to right.

%
   \begin{figure*}[t]
       \centering
\plottwo{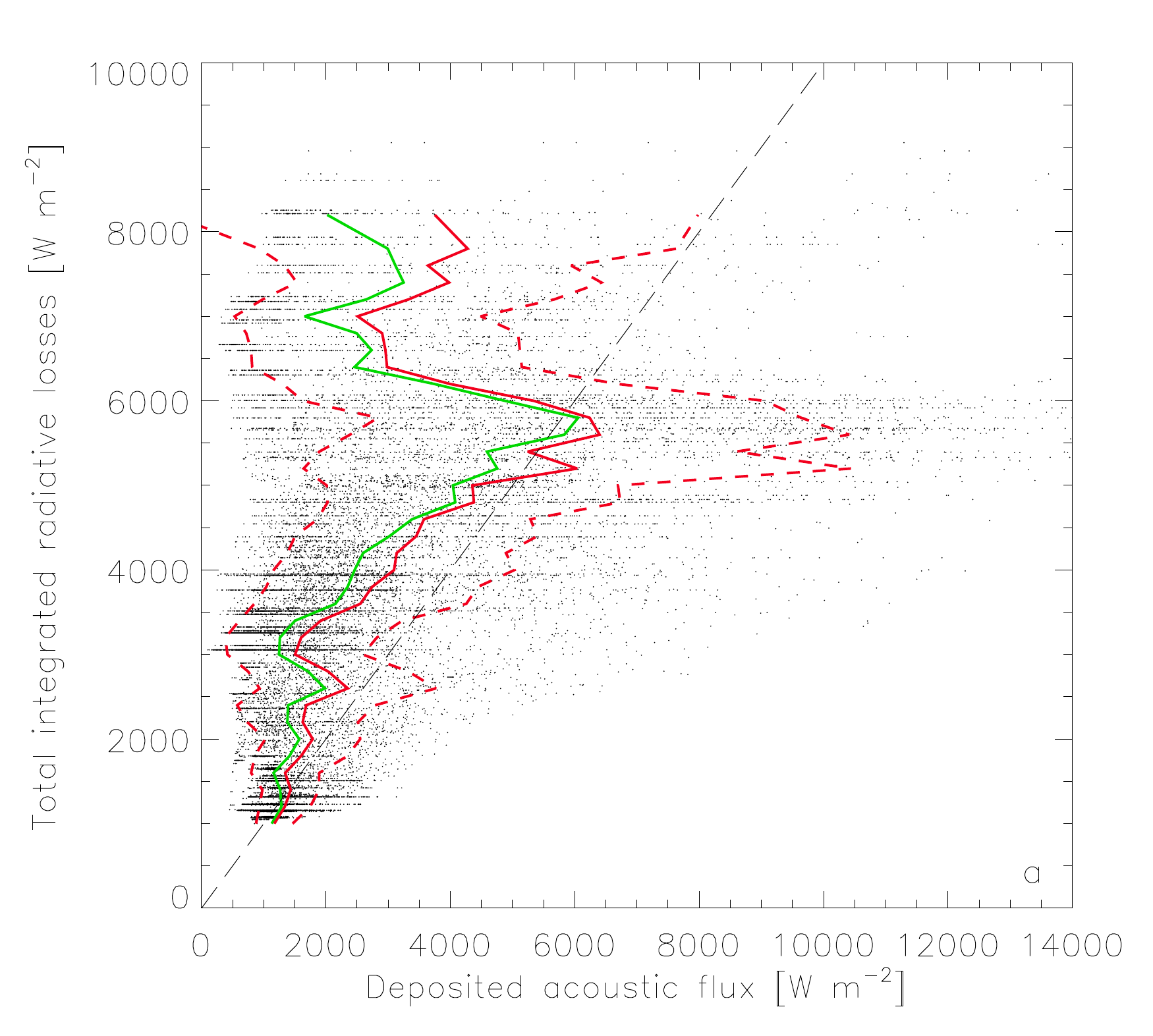}{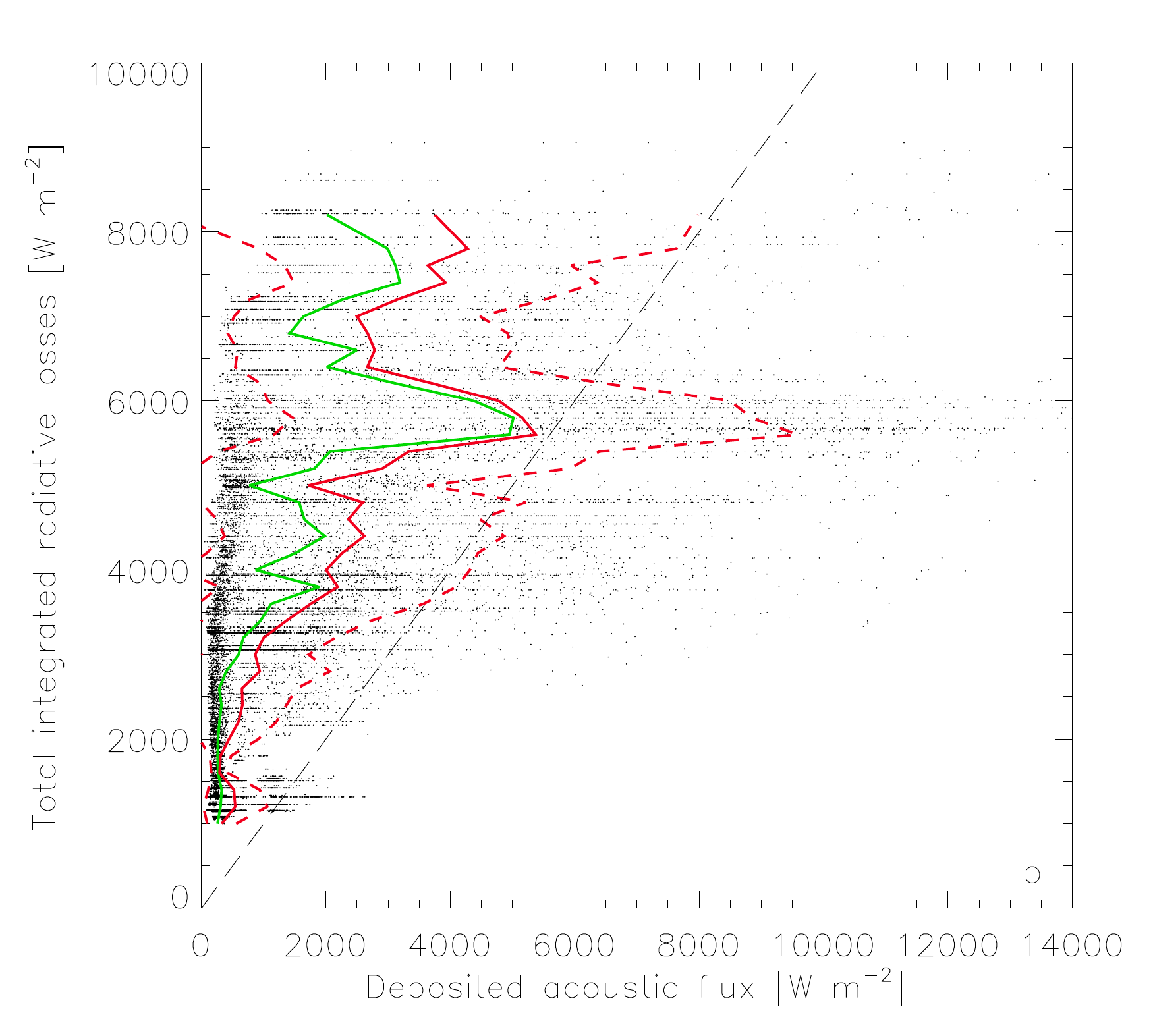}
\caption{Scatter plots of total integrated radiative losses versus deposited acoustic flux
in the region of interest (19516  points). The acoustic fluxes are calculated using (a) the
original magnetic inclination angles and (b) the corrected ones.
Solid lines show average (red) and median (green) values together with red dashed lines
of $\pm 1\sigma$. The straight dashed line represents the full balance of radiative
losses by acoustic flux deposit.}
       \label{Fig:srov}
   \end{figure*}
%

The total net radiative cooling rates (a sum of the \ion{Ca}{2}, \ion{Mg}{2}, and H
contributions) are calculated for each model in ROI and integrated over the height range
900--1500 km. The right panel of Figure~\ref{Fig:T-example} shows examples of the
height dependence of cooling rates in the quiet Sun and plage at the same positions
as of the temperature stratifications. In the plage, the total cooling rates have a maximum
at $h = 1000$--1100~km, which corresponds to maximum cooling rates of \ion{Ca}{2}
(the \ion{Mg}{2} contribution is by an order lower).
A secondary peak, caused by enhanced contribution of H and H$^-$ continua in hot
and dense atmospheres, appears around $h = 800$~km in the brightest part of the plage
(orange curves). It does not enter our integration range.
The map of the total integrated radiative losses is depicted in Figure~\ref{Fig:rlosses}a.

We calculate the deposited acoustic fluxes (see Section~\ref{sec:obs}) using
two different sets of magnetic field inclinations to estimate the influence of uncertainties
in the inclination angle:
(i) Original inclination angles retrieved from the inversion, including the unreliable values
near 90\degr ~in weak-field or non-magnetic regions. The resulting  map
of the deposited acoustic flux is shown in Figure~\ref{Fig:rlosses}b.
(ii) Corrected inclination angles, where the unreliable values are set to zero by means of the
mask described in Section~\ref{sec:obs}. The corresponding deposited acoustic flux map is
shown in in Figure~\ref{Fig:rlosses}c.
The coefficient of spatial correlation between the total integrated radiative losses and the
deposited acoustic flux is 0.47 in the case of (i) and 0.50 in the case of (ii).

Scatter plots of total integrated radiative losses $L$ versus the acoustic energy flux
$\Delta F_{\rm ac}$ deposited in the chromosphere between $h = 900$ and 1500 km
are shown in Figure~\ref{Fig:srov}a and b for the original inclination angles and the
corrected ones, respectively.
The red solid line represents mean values of $\Delta F_{\rm ac}$ that fall into
200 W m$^{-2}$ wide bins of the $L$ histogram. The bins must contain at least
100 points to calculate the mean value. Likewise, the green line represents median values.
The red dashed lines delimit the $\pm 1\sigma$ range that characterizes the scatter of
individual points in each bin.
When the unreliable horizontal magnetic inclination is included (Figure~\ref{Fig:srov}a),
the deposited acoustic fluxes in the quiet-Sun area  ($L < 5000$ W m$^{-2}$)
almost double those calculated using the corrected
inclination values (Figure~\ref{Fig:srov}b). This is not realistic because of too many
points where $\Delta F_{\rm ac} > L$. The increase of $\Delta F_{\rm ac}$ is much
smaller in the plage, where the magnetic inclination is determined reliably. In the
further discussion, we shall use the deposited acoustic fluxes calculated using the
corrected inclination angles.

For most of the points (88 \%) in the plot (Figure~\ref{Fig:srov}b),
$L > \Delta F_{\rm ac}$, that is, the deposited acoustic flux is insufficient to balance
the radiative losses and maintain the (semi-empirical) temperature at corresponding
positions in ROI.
The scatter of $\Delta F_{\rm ac}$ versus $L$ is large, so that we have to
express the contribution of the deposited acoustic flux to the radiative losses statistically,
using the ratios $\overline{\Delta F_{\rm ac}} / \overline{L}$ and
${\rm median} (\Delta F_{\rm ac}) / \overline{L}$. Their values start at 0.3 (median 0.2)
for $L < 3500$ W~m$^{-2}$ and increase to 0.5 (median 0.3) for
$3500 < L <5000$ W~m$^{-2}$ in the quiet area.
They reach 0.7 (median 0.6) with a peak of 0.9 for $5000 < L <6500$ W~m$^{-2}$
at the periphery of the plage, where the magnetic inclination is large, and drop to
0.5 (median 0.4) for $L >6500$ W~m$^{-2}$ in the brightest part of the plage with
the nearly vertical magnetic field.
The points with $L < \Delta F_{\rm ac}$ will be discussed in Section~\ref{sec:disc}.

Compared to the results of of \citetalias{Sobotka16}, the median values of
$\Delta F_{\rm ac}$ are consistent excepting the brightest parts of the plage
($L > 6500$ W~m$^{-2}$), where the magnetic inclination was overestimated.
Moreover, the revision of the magnetic inclination map resulted in an increase of
$\Delta F_{\rm ac}$ at the plage periphery ($L \simeq 6000$ W~m$^{-2}$).
Other differences can be explained by improved values of the gas density, to which the
acoustic fluxes are directly proportional. The present set of 1737 atmospheric models
provides a more realistic density distribution than the simple grid of only seven models.

\section{Discussion and conclusions} \label{sec:disc}

A quantitative comparison of deposited acoustic energy flux with total integrated
radiative losses in the middle chromosphere of the quiet Sun and a weak plage is made.
The comparison is based on a consistent set of high-resolution observations acquired
by the IBIS instrument in the line \ion{Ca}{2} 854.2~nm. The deposited acoustic
flux is derived from Doppler velocities observed in the line core and a set of 1737
non-LTE 1D hydrostatic semi-empirical models, which are also used for the calculation
of radiative losses. The models are obtained by scaling the temperature and column
mass of five initial models VAL B--F \citep{Vernazza81} to get the best fit of
synthetic to time-averaged observed profiles of the \ion{Ca}{2} 854.2~nm line.

\begin{figure}[t]
       \centering
\includegraphics[width=0.5\textwidth]{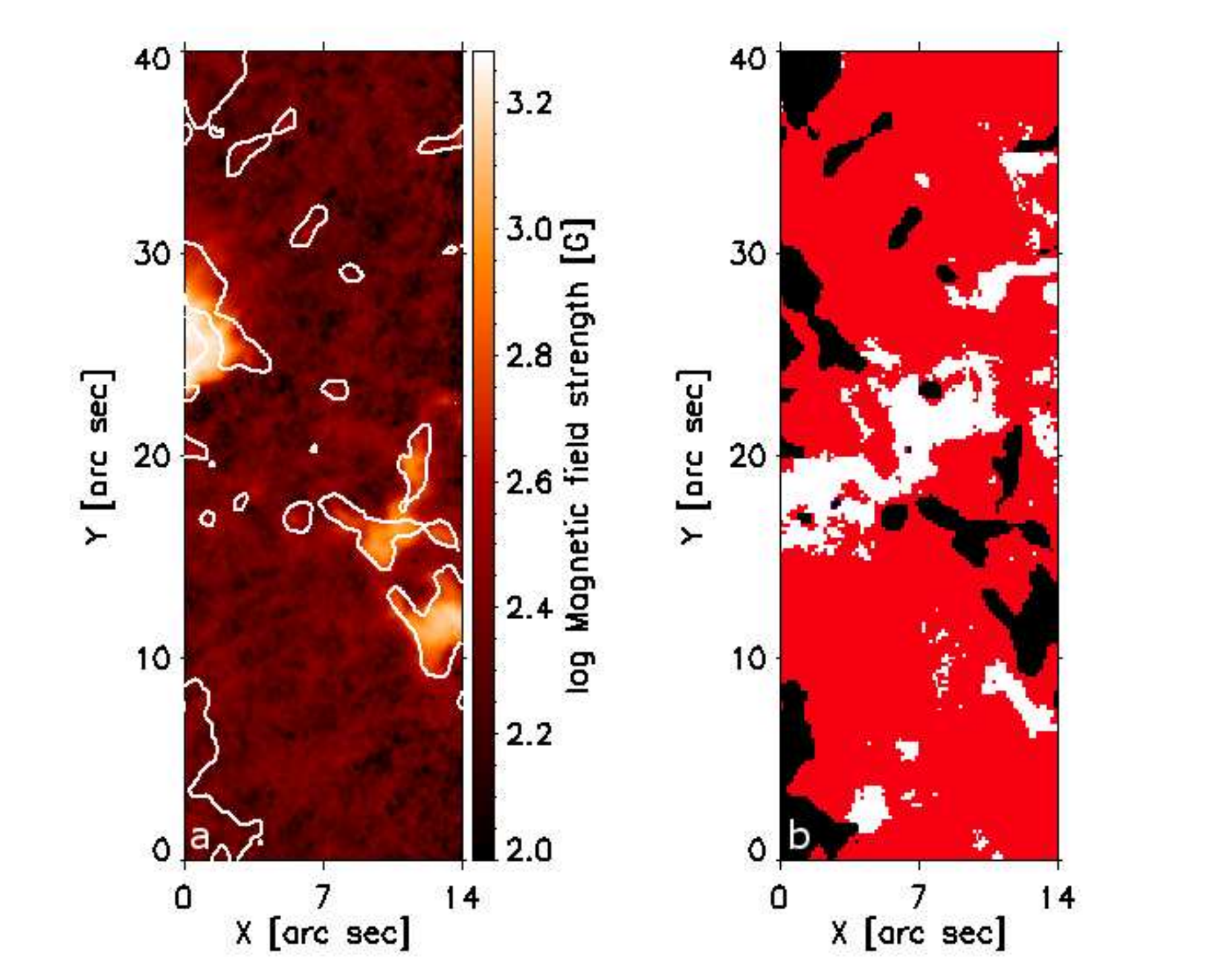}
\caption{(a) Map of magnetic field strength (in logarithmic scale) with contours of areas
where the fit of synthetic to observed \ion{Ca}{2} line profiles is worse than $2\sigma$.
(b) Areas of radiative losses smaller than the deposited acoustic flux (white)
together with the bad-fit areas (black).}
       \label{Fig:masks}
   \end{figure}

The fit quality $\chi^2_{\rm min}$ changes with the position in ROI. We define
the bad-fit areas where $\chi^2_{\rm min}$ is larger than $2\sigma$ of its statistical
distribution. Several of them, where the observed profiles are broader than the
synthetic ones, coincide with regions of enhanced magnetic field in the pore and plage.
Contours that outline the bad-fit areas together with the map of magnetic-field strength
are shown in Figure~\ref{Fig:masks}a. The Zeeman broadening of the \ion{Ca}{2}
line with Land\'e factor $g = 1.1$ probably does not play a major role. Rather,
this effect might be explained by the fact that our 1D semi-empirical models do not
account for the complex 3D distribution of thermodynamic and magnetic parameters in
such areas.
The bad-fit areas represent 12.5 \% ~of the data set and a mask is made to remove
the affected points from the scatter plot of $L$ versus $\Delta F_{\rm ac}$.
This removal alters the plot shown in Figure~\ref{Fig:srov}b and the derived
statistical values only in the part of $L > 6500$ W~m$^{-2}$, corresponding to the
brightest plage region, where $\overline{\Delta F_{\rm ac}}$ becomes larger by
the factor of 1.3 on average and its contribution to the total integrated radiative
losses $\overline{\Delta F_{\rm ac}} / \overline{L}$ increases from 0.5 to 0.6.

%
%

In 12 \% ~points of our data set, the deposited acoustic energy flux is larger
than the total integrated radiative losses ($L < \Delta F_{\rm ac}$).
Regions formed by these points are shown in Figure~\ref{Fig:masks}b together with the
bad-fit ones in white and black colors, respectively.
The $L < \Delta F_{\rm ac}$ regions appear at locations where
{\bf the magnetic inclination is larger than 60\degr ~and}
temporal variations of Doppler velocity and intensity of the \ion{Ca}{2} 854.2 nm core are strong
(cf. Figure~1 and its animation in \citetalias{Sobotka16}): around the border between the
pore's superpenumbra and plage (subfield 0\arcsec$<x<$10\arcsec, 15\arcsec$<y<$25\arcsec)
and partly at the periphery of the plage.

We have shown that the deposited acoustic-flux energy in the quiet-Sun chromosphere
balances 30--50 \% of the energy released by radiative losses. The energy carried by
(magneto)acoustic waves in the plage supplies 50--60 \% of the radiated energy at locations
with vertical magnetic field and 70--90 \% in regions where the magnetic field is inclined more
than 50\degr. These values are statistical averages of results with a large individual scatter and
they are based on one observation of a single small solar area. They are also critically
sensitive to the correct determination of the magnetic field inclination, particularly in the
quiet-Sun region.
We also have to note that the area considered as quiet Sun in our ROI is close to the plage
and it  falls within the extended canopy region of the plage's magnetic field (see Figure~1
in \citetalias{Sobotka16}).  The effect of ``magnetic shadows'', which are related to the
elevated magnetic field forming the \ion{Ca}{2} fibrils \citep{Vecchio07}, reduces
the oscillatory power in this region, as seen from the power maps in Figure~7 of
\citet{Sobotka13}. Consequently, the deposited acoustic flux in our ``quiet area'' may be
lower than that in quiet regions far from the plage. Additional studies of different quiet
and active regions in various chromospheric lines together with precise measurements of
magnetic field inclination are needed to obtain more general and conclusive results.

\newpage
ACKNOWLEDGMENTS \\

We thank the anonymous reviewer for valuable comments and suggestions.
This work was supported by the Czech Science Foundation and Deutsche Forschungsgemeinschaft
under the common grant 18-08097J -- DE 787/5-1 and the institutional support RVO:67985815
of the Czech Academy of Sciences.
The Dunn Solar Telescope was run by the National Solar Observatory (NSO), which is operated
by the Association of Universities for Research in Astronomy, Inc. (AURA), for the National
Science Foundation.
IBIS has been built by INAF/Osservatorio Astrofisico di Arcetri with contributions from the Universities
of Firenze and Roma ``Tor Vergata'', NSO, and the Italian Ministries of Research (MIUR) and
Foreign Affairs (MAE). The observations at IBIS were supported by the Rome Tor Vergata
``Innovative techniques and technologies for the study of the solar magnetism'' grant funded by MIUR.

\bibliography{Chromheat19s2}{}
\bibliographystyle{aasjournal}

\end{document}